\documentclass{article}
\usepackage{alife10}

\title{Evolution of Complexity:\\
Introduction to the Workshop}
\author{Carlos Gershenson$^1$ and Tom Lenaerts$^2$\\
\mbox{}
$^1$ CLEA, ECCO, Vrije Universiteit Brussel, Brussels, Belgium\\
\mbox{}
$^2$ SWITCH, Flanders Interuniversity Institute for Biotechnology, Vrije Universiteit Brussel Brussels, Belgium\\
$\{$ cgershen $|$  tlenaert$\}$ @vub.ac.be}

\begin{document}
\maketitle


The evolution of complexity has been a central theme for Biology and Artificial Life \cite{Bonner1988,problems}. Complexification has been interpreted in different ways: as a process of diversification between evolving units \cite{Bonner1988} or as a scaling process that is related to the idea of transitions between different levels of complexity \cite{szathmary97}. There have been previous workshops on this topic, e.g. \cite{HeylighenEtAl1999,wdh02a}, but many open questions still remain. 
Trying to answer these questions from a general perspective might not immediately produce concrete answers for the different fields that are interested in this topic.  Consequently, this workshop is organised with a particular focus in mind: \emph{The emergence of complexity through evolutionary mechanisms.}
Primarily we want to have a discussion on evolutionary and related dynamics as mechanisms for producing complexity. Furthermore, we want to bring together historical and novel research in this context.

In the call for papers, the following list of questions were posed, which might  be addressed at the workshop:
\begin{itemize}
\item What are the environmental constraints of complexity growth in living systems?
\item  What is the origin and role of developmental mechanisms in evolution?
\item  Are the principles of natural selection, as they are currently understood, sufficient to explain the evolution of complexity?
\item  What are the limits at different levels to the evolution of complexity, and which conditions could reduce evolved complexity?
\item  Which model/terminology is more appropriate to  understand/speak about the evolution of complexity in living systems?
\item  How could complexity growth be measured or operationalised in natural and artificial systems?
\item  How can data from nature be brought to bear on the study of this issue?
\item  What are the main hypotheses about complexity growth that can actually be tested?
\item  Is it possible to direct/manipulate the evolution of complexity, or which benefits would bring its understanding?
\end{itemize}

In answer to these question, we received a number of contributions by different researchers, whose manuscripts follow immediately after this introduction.  The articles are concerned with ways to generate, measure and formalise complexity.  Moreover, due to the long history concerning this topic, all articles provide meaningful pointers to earlier work on this topic.  Given all this information we hope that this workshop will provide the incentive for other Artificial Life researchers to try to provide answers to the previous questions. 

A set of online discussion forums has been created at the workshop's website\footnote{http://ecco.vub.ac.be/ECO/} to promote the discussion before and after the workshop. The reader is kindly invited to read and participate in these forums.

\section{Acknowledgements}

We would like to thank the entire programme committee members (Lee Altenberg, Mark Bedau, Hugues Bersini, John Bonner, Dominique Chu, Jim Crutchfield, Bruce Edmonds, Carlos Gershenson, Mario Giacobini, Franics Heylighen, Tom Lenaerts, Juan Juli\'{a}n Merelo, Barry McMullin, Melanie Mitchell, Chrystopher Nehaniv, Jorge Pacheco, Tom Ray, Jon Rowe, Stanley Salthe, Cosma Shalizi, Peter Schuster, E\"{o}rs Szathm\'{a}ry, and Richard Watson) for their reviews and suggestions.  We thank all researchers who send in an article for this workshop. Furthermore, we highly appreciate the willingness of Jim Crutchfield and Mark Bedau to give a more lengthy discussion, as invited speakers, on their perspectives on the evolution of complexity.  Finally we want to thank the organisers of Alife X for hosting this workshop.

\bibliography{ECO-WS-Intro}
\bibliographystyle{alife10}

\end{document}